\documentclass[12pt]{iopart}
\usepackage{bbm}
\usepackage{graphicx}
\usepackage{iopams}
\newcommand{\ket}[1]{\vert #1 \rangle}

\newcommand{\dket}[1]{\vert #1 \rangle\rangle}

\newcommand{\calpha}{\alpha^*}
\newcommand{\cbeta}{\beta^*}
\newcommand{\cxi}{\xi^*}

\newcommand{\wtA}{\widetilde{A}}
\newcommand{\wtB}{\widetilde{B}}
\newcommand{\bmsigma}{\boldsymbol \sigma}
\newcommand{\bmX}{\boldsymbol X}
\newcommand{\bbGamma}{{\rm I}\!\Gamma}
\newcommand{\calN}{{\cal N}}
\begin{document}
\title{Photon subtracted states and enhancement of nonlocality
in the presence of noise}
\date{\today}
\author{Stefano Olivares\footnote{Stefano.Olivares@mi.infn.it} 
and Matteo G. A. Paris\footnote{Matteo.Paris@fisica.unimi.it}}
\address{Dipartimento di
Fisica and INFM, Universit\`a degli Studi di Milano, Italia}
\begin{abstract}
We address nonlocality of continuous variable systems in the presence 
of dissipation and noise. Three nonlocality tests have been considered, 
based on the measurement of displaced-parity, field-quadrature and 
pseudospin-operator, respectively. Nonlocality of twin-beam has been 
investigated, as well as that of its non-Gaussian counterparts obtained 
by inconclusive subtraction of photons. Our results indicate that: 
i) nonlocality of twin-beam is degraded but not destroyed by noise; 
ii) photon subtraction enhances nonlocality in the presence of noise, 
especially in the low-energy regime.
\end{abstract}
\section{Introduction}\label{s:1}
Nonlocality, {\em i.e.} the existence of correlations which cannot 
be explained by any local hidden variable model, is perhaps 
the most debated implication of quantum mechanics. During the last 
decade other aspects of nonlocality, in addition to
generating nonlocal correlations, have been discovered. For example, the
possibility of teleporting and effectively encoding information, 
as well as the ability to perform certain computations exponentially 
faster than any classical device.
\par
Realistic  implementations of quantum information protocols
require the investigation of nonlocality properties of quantum 
states in a noisy environment. In particular, the robustness of 
nonlocality should be addressed, as well as the design of protocols 
to preserve and possibly enhance nonlocality in the presence of noise.
\par
The evolution of nonlocality for a twin-beam state of radiation (TWB) in a
thermal environment was studied in Ref.~\cite{jeong:noise} by means of the
displaced parity test \cite{bana}, whereas in Ref.~\cite{filip:PRA:66} its
nonlocality was investigated using the pseudospin operators
\cite{chen:PRL:88} when only dissipation occurs.
\par
In Ref.~\cite{ips:PRA:67} we have suggested a conditional measurement
scheme on TWB leading to a non-Gaussian entangled mixed state, which
improves fidelity of teleportation of coherent states. This process,
termed inconclusive photon subtraction (IPS), is based on mixing each mode
of the TWB with the vacuum in an unbalanced beam splitter and then
performing inconclusive photodetection on both modes, {\em i.e.} revealing
the reflected beams without discriminating the number of the detected
photons. IPS states have the following properties: they improve the
teleportation fidelity for coherent states \cite{ips:PRA:67} and show
enhanced nonlocal correlations in the phase space \cite{ips:PRA:70} in
ideal conditions, namely in the absence of noise. Motivated by these
results and by the recent experimental generation of IPS states
\cite{wenger:PRL:04}, in this paper we extend the previous studies on the
TWB and consider the nonlocality of the IPS state in the presence of noise.
\par
The paper is structured as follows. In Sec.~\ref{s:lossy} we address the
evolution of the TWB in a noisy channel where both dissipation and thermal
noise are present, whereas in Sec.~\ref{s:IPS} we briefly review the IPS
process. In Secs.~\ref{s:DP}, \ref{s:HD} and \ref{s:PS} we investigate the
nonlocality of TWB and IPS by means of three different tests: displaced
parity, homodyne detection, and pseudospin test, respectively. Finally,
Sec.~\ref{s:remarks} closes the paper with some concluding remarks.
\section{Dynamics of TWB in noisy channels} \label{s:lossy}
The so called twin-beam state of radiation (TWB), {\em i.e.}
$\dket{\Lambda} = \sqrt{1-\lambda^2}\sum_k \lambda^2 \ket{k}\otimes\ket{k}$
with $\lambda=\tanh r$, $r$ being the TWB squeezing parameter.
$\dket{\Lambda}$ is obtained by parametric down-conversion of the vacuum,
$\dket{\Lambda} = \exp\{ r(a^\dag b^\dag - ab) \}\ket{0}$, $a$ and $b$
being field operators, and it is described by the Gaussian Wigner function
\begin{equation}
W_{0}(\alpha,\beta) =
\frac{\exp\{
-2 \widetilde{A}_0 (|\alpha|^2+|\beta|^2)
+ 2 \widetilde{B}_0 (\alpha\beta + \calpha\cbeta) \}}
{4\pi^2\sqrt{{\rm Det}[\bmsigma_0]}}\,,
\label{twb:wig}
\end{equation}
with
\begin{equation}
\widetilde{A}_0 = \frac{A_0}{16 \sqrt{{\rm Det}[\bmsigma_0]}}\,,\qquad
\widetilde{B}_0 = \frac{B_0}{16 \sqrt{{\rm Det}[\bmsigma_0]}}\,,
\end{equation}
where $A_0 \equiv A_0(r) = \cosh(2 r)$,
$B_0 \equiv B_0(r) = \sinh (2 r)$ and $\bmsigma_0$ is the covariance matrix
\begin{equation}\label{cvm:twb}
\bmsigma_0 = \frac14
\left(
\begin{array}{cc}
A_0\, \mathbbm{1}_2 & B_0\, \bmsigma_3\\[1ex]
B_0\, \bmsigma_3 & A_0\, \mathbbm{1}_2
\end{array}\right)\:,
\end{equation}
$\mathbbm{1}_2$ being the $2 \times 2$ identity matrix and $\bmsigma_3 =
{\rm Diag}(1,-1)$. Using a more compact form, Eq.~(\ref{twb:wig}) can
also be rewritten as
\begin{equation}\label{gauss:form}
W_{0}(\bmX) =
\frac{\exp\left\{ -\frac12\, \bmX^{T}\,\bmsigma_{0}^{-1}\,\bmX \right\}}
{4 \pi^2 \sqrt{{\rm Det}[\bmsigma_0]}}\,,
\end{equation}
with $\bmX = (x_1,y_1,x_2,y_2)^{T}$, $\alpha=x_1+iy_1$ and
$\beta=x_2+iy_2$, and $(\cdots)^{T}$ denoting the transposition operation.
\par
When the two modes of the TWB interact with a noisy environment, namely in the
presence of dissipation and thermal noise, the evolution of the Wigner
function (\ref{twb:wig}) is described by the following Fokker-Planck equation
\cite{wm:quantopt:94,binary,seraf:PRA:69}
\begin{equation}\label{fp:eq:cmp}
\partial_t W_{t}(\bmX) = \frac12 \Big(
\partial_{\bmX}^T \bbGamma \bmX + \partial_{\bmX}^T
\bbGamma \bmsigma_{\infty} \partial_{\bmX} \Big) W_{t}(\bmX)\,,
\end{equation}
with $\partial_{\bmX} =
(\partial_{x_1},\partial_{y_1},\partial_{x_2},\partial_{y_2})^{T}$.
The damping matrix is given by $\bbGamma = \bigoplus_{k=1}^2\,
\Gamma_k \mathbbm{1}_2$, whereas
\begin{eqnarray}
\bmsigma_{\infty} &= \bigoplus_{k=1}^{2}\, \bmsigma_{\infty}^{(k)} =
\left(
\begin{array}{cc}
\bmsigma_{\infty}^{(1)} & \boldsymbol{0} \\[1ex]
\boldsymbol{0} & \bmsigma_{\infty}^{(2)}
\end{array}
\right)\,,
\end{eqnarray}
where $\boldsymbol{0}$ is the $2 \times 2$ null matrix and
\begin{equation}
\bmsigma_{\infty}^{(k)} = 
\frac14
\left(
\begin{array}{cc}
1 + 2 N_{k} & 0\\[1ex]
0 & 1 + 2 N_k
\end{array}
\right)\,.
\end{equation}
$\Gamma_k$, $N_k$ denotes the damping rate and the average number of
thermal photons of the channel $k$, respectively. $\bmsigma_{\infty}$
represents the covariance matrix of the environment and, in turn, the
asymptotic covariance matrix of the evolved TWB.  Since the environment is
itself excited in a Gaussian state, the evolution induced by
(\ref{fp:eq:cmp}) preserves the Gaussian form (\ref{gauss:form}).  The
covariance matrix at time $t$ reads as follows
\cite{seraf:PRA:69,FOP:napoli:05}
\begin{equation}
\bmsigma_t = \mathbbm{G}_t^{1/2}\,\bmsigma_0\,\mathbbm{G}_t^{1/2}
+ (\mathbbm{1} - \mathbbm{G}_t)\,\bmsigma_{\infty}\,,
\end{equation}
where $\mathbbm{G}_t = \bigoplus_{k=1}^2\,e^{-\Gamma_k t}\,\mathbbm{1}_2$.
The covariance matrix $\bmsigma_t$ can be also written as
\begin{equation}\label{evol:cvm:12}
\bmsigma_t = \frac 14
\left(
\begin{array}{cc}
A_t(\Gamma_1,N_1)\, \mathbbm{1}_2&  B_t(\Gamma_1)\,\bmsigma_3 \\[1ex]
B_t(\Gamma_2)\, \bmsigma_3 & A_t(\Gamma_2,N_2)\, \mathbbm{1}_2 
\end{array}
\right)
\end{equation}
with
\begin{equation}
\label{AtBt}
\eqalign{
&A_t(\Gamma_k,N_k) = A_0\,e^{-\Gamma_k t}
+ \left(1-e^{-\Gamma_k t}\right) (1 + 2 N_k)\,,\\  
&B_t(\Gamma_k) = B_0\,e^{-\Gamma_k t}\,.
}
\end{equation}
\par
Let us now consider channels with the same damping rate $\Gamma$ but
different number of thermal photons, $N_1$ and $N_2$: using the density
matrix formalism, the state corresponding to the covariance matrix
(\ref{evol:cvm:12}) has the following form
\begin{equation}\label{rho:t:evol}
\varrho_t = S_2(\xi)\,\mu_1\otimes\mu_2\,S_2^{\dag}(\xi)\,,
\end{equation}
where $\mu_k$ is the thermal state
\begin{equation}
\mu_k = \frac{1}{1 + M_k}
\left( \frac{M_k}{1+M_k} \right)^{a^{\dag}_k a_k}
\end{equation}
$a_k$, $k=1,2$ being the mode operators. The average number of photons are
given by
\begin{eqnarray} 
M_1 &=
\frac14 \left[ \sqrt{A_{+}^2 - 16 B_t} -
(2 - A_{-}) \right]\label{m1:evol}\,,\\
M_2 &=
\frac14 \left[ \sqrt{A_{+}^2 - 16 B_t} -
(2 + A_{-}) \right]\label{m2:evol}\,,
\end{eqnarray}
$A_{\pm} = A_{1,t} \pm A_{2,t}$, $A_{k,t}\equiv A_{t}(\Gamma,N_k)$ and
$B_t=B_t(\Gamma)$. In Eq.~(\ref{rho:t:evol}) $S_2(\xi)= \exp\{ \xi
a_1^{\dag}a_2^{\dag} - \cxi a_1 a_2 \}$ denotes the two-mode squeezing
operator, with parameter $\xi \in \mathbb{C}$
\begin{eqnarray}
&|\xi| = \sinh^{-1}\left( \sqrt{
\frac{A_{+}}
{2(A_{+}^2 - 16 B_t)^{1/2}}-\frac12}\right)\label{x1:evol}\,,\\
&\arg[\xi] = \pi/2\label{arg:xi:evol}\,.
\end{eqnarray}
Eq.~(\ref{rho:t:evol}) says that the quantum state of a TWB, after
propagating in a noisy channel, is the same of a state obtained by
parametric down-conversion from a noisy background \cite{FOP:napoli:05}.
Their properties, and in particular entanglement and nonlocality, can be
addressed in an unified way using Eq.~(\ref{rho:t:evol}) or, equivalently,
Eqs.~(\ref{evol:cvm:12}) and (\ref{AtBt}).
\par
Finally, if we assume $\Gamma_1 = \Gamma_2 = \Gamma$ and $N_1 = N_2 = N$,
then the covariance matrix (\ref{evol:cvm:12}) becomes formally identical
to (\ref{cvm:twb}) and the corresponding Wigner function reads
\begin{equation}
W_{t}(\alpha,\beta) =
\frac{\exp\{
-2 \widetilde{A}_t (|\alpha|^2+|\beta|^2)
+ 2 \widetilde{B}_t (\alpha\beta + \calpha\cbeta)\}}
{4\pi^2\sqrt{{\rm Det}[\bmsigma_t]}}\,,
\label{twb:wig:noise}
\end{equation}
with
\begin{equation}
\widetilde{A}_t = \frac{A_t(\Gamma,N)}{16\sqrt{{\rm Det}[\bmsigma_t]}}\,,
\qquad  
\widetilde{B}_t = \frac{B_t(\Gamma)}{16\sqrt{{\rm Det}[\bmsigma_t]}}\,,
\end{equation}
whereas the density matrix, {\em mutatis mutandis}, is still given by
Eq.~(\ref{rho:t:evol}).
\section{De-Gaussification and noise}\label{s:IPS}
\begin{figure}
\begin{center}
\includegraphics[scale=.8]{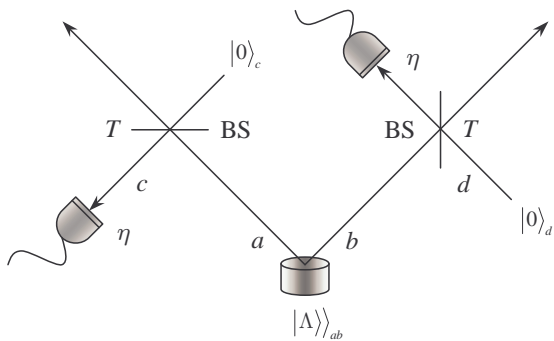}
\end{center}
\vspace{-.3cm}
\caption{\label{f:IPS:scheme} Scheme of the IPS process.}
\end{figure}
When thermal noise and dissipation affect the propagation of an entangled
state, its nonlocal properties are reduced and, finally, destroyed
\cite{binary,seraf:PRA:69,rossi:JMO:04}. Therefore it is of interest to
look for some technique in order to preserve, at least in part, such
correlations, or to enhance the nonlocality of the state which will face
the lossy transmission line. Since it has been shown that the
de-Gaussification of a TWB can enhance its entanglement in the ideal case
and since non-Gaussian states can be produced using the current technology
\cite{wenger:PRL:04}, in this and the following Sections we will
investigate whether or not this process can be useful also in the presence
of noise.
\par
The de-Gaussification of a TWB can be achieved by subtracting photons from
both modes \cite{ips:PRA:67,opatr:PRA:61,coch:PRA:65}.  In
Ref.~\cite{ips:PRA:67} we referred to this process as to inconclusive
photon subtraction (IPS) and showed that the resulting state, the IPS
state, can be used to enhance the teleportation fidelity of coherent states
for a wide range of the experimental parameters. Moreover, in
Ref.~\cite{ips:PRA:70}, we have shown that, in the absence of any noise
during the transmission stage, the IPS state has nonlocal correlations
larger than those of the TWB irrespective of the IPS quantum efficiency
(see also Refs.~\cite{nha:PRL:93,garcia:PRL:93}).
\par
First of all we briefly recall the IPS process, whose scheme is sketched in
Fig.~\ref{f:IPS:scheme}. The two modes, $a$ and $b$, of the TWB are mixed
with the vacuum (modes $c$ and $d$, respectively) at two unbalanced beam
splitters (BS) with equal transmissivity; the  modes $c$ and $d$ are then
revealed by avalanche photodetectors (APDs) with equal efficiency, which
can only discriminate the presence of radiation from the vacuum: the IPS
state is obtained when the two detectors jointly click. The mixing with the
vacuum at a beam splitter with transmissivity $T$ followed by the on/off
detection with quantum efficiency $\eta$ is equivalent to mixing with an
effective transmissivity $\tau$ \cite{ips:PRA:67}
\begin{equation}
\tau \equiv \tau(T,\eta) = 1 - \eta (1-T)\,,
\end{equation}
followed by an ideal ({\em i.e.} efficiency equal to $1$) on/off detection.
Using the Wigner formalism, when the input state arriving at the two beam
splitters is the TWB $W_{0}(\alpha,\beta)$ of Eq.~(\ref{twb:wig}), the
state produced by the IPS process reads as follows (see
Ref.~\cite{ips:PRA:70} for the details about the calculation and about the
de-Gaussification map for the density matrix and Wigner function in the
case of a TWB)
\begin{equation}\label{ips:wigner}
W_{0}^{\rm (IPS)}(\alpha,\beta) =
\frac{1}{\pi^2\,p_{11}(r,\tau)} 
\sum_{k=1}^4 {\cal C}_k(r,\tau)\,
W_{r,\tau}^{(k)}(\alpha,\beta)\,,
\end{equation}
where
\begin{equation}\label{ips:probability}
p_{11}(r,\tau) = \sum_{k=1}^4 \frac{{\cal C}_k(r,\tau)}{
(b-f_k)(b-g_k)-(2 \wtB_0 \tau + h_k)^2}\,
\end{equation}
is the probability of a click in both the APDs. In Eqs.~(\ref{ips:wigner})
and (\ref{ips:probability}) we introduced
\begin{equation}
\fl
W_{r,\tau}^{(k)}(\alpha,\beta) =
\exp\{ -(b-f_k) |\alpha|^2 -(b-g_k) |\beta|^2
+ (2 \wtB_0 \tau + h_k) (\alpha\beta + \calpha\cbeta)\}\,,
\end{equation}
and defined
\begin{equation}
{\cal C}_k(r,\tau)=
\frac{C_k}
{ \sqrt{{\rm Det}[\bmsigma_0]}\,[x_k y_k - 4 \wtB_0^2 (1-\tau)^2]}\,,
\end{equation}
where $C_1 = 1$, $C_2 = C_3 = -2$, $C_4 = 4$; $x_k \equiv x_k(r,\tau)$, and
$y_k \equiv y_k(r,\tau)$ are
\begin{eqnarray}
&x_1 = x_3 = y_1 = y_1 = a \nonumber\\
&x_2 = x_4 = y_3 = y_4 = a +2 \nonumber
\end{eqnarray}
with $a \equiv a(r,\tau) = 2 [\wtA_0 (1-\tau) +
\tau]$,
$b \equiv b(r,\tau) = 2 [\wtA_0 \tau + (1-\tau)]$; finally, $f_k$, $g_k$,
and $h_k$ depend on $r$ and $\tau$ and are given by
\begin{eqnarray}
f_k & = \calN_k
\, [x_k \wtB_0^2 + 4 \wtB_0^2 (1-\wtA_0) (1-\tau) + y_k (1-\wtA_0)^2]\,,\\
g_k &= \calN_k
\, [x_k (1-\wtA_0)^2 + 4 \wtB_0^2 (1-\wtA_0) (1-\tau) + y_k \wtB_0^2]\,,\\
h_k &= \calN_k
\, \{(x_k + y_k) \wtB_0 (1-\wtA_0) + 2 \wtB_0 [\wtB_0^2 + (1-\wtA_0)^2]
(1-\tau)\}\,,\\
\calN_k &\equiv \calN_k(r,\tau) = {\displaystyle
\frac{4 \tau\, (1-\tau)}{x_k y_k - 4 \wtB_0^2
(1-\tau)^2}\,.
}
\end{eqnarray}
The state corresponding to Eq.~(\ref{ips:wigner}) is no longer a Gaussian
state and its nonlocal properties, in ideal conditions, were studied in
Ref.~\cite{ips:PRA:70}.
\par
Here we are interested in the case when the IPS process is performed on a
TWB evolved in a noisy environment with both the channels having the same
damping rate and thermal noise. The Wigner function of the state arriving
at the beam splitters is now given by Eq.~(\ref{twb:wig:noise}), and the
output state is still described by Eq.~(\ref{ips:wigner}), but with the
following substitutions
\begin{equation}\label{sostituzioni}
\wtA_0 \to \widetilde{A}_t \,,\quad
\wtB_0 \to \widetilde{B}_t \,,\quad
\bmsigma_0 \to \bmsigma_t\,.
\end{equation}
We will denote with $W_{\Gamma,N}^{\rm (IPS)}(\alpha,\beta)$ the Wigner
function of this degraded IPS state.
\par
In the next Sections we will analyze the nonlocality of the IPS state in
the presence of noise by means of Bell's inequalities.
\section{Nonlocality in the phase space} \label{s:DP}
Parity is a dichotomic variable and thus can be used to establish
Bell-like inequalities \cite{CHSH}. 
The displaced parity operator on two modes is defined as \cite{bana}
\begin{equation}
\hat{\Pi}(\alpha,\beta) =
D_a(\alpha)(-1)^{a^\dag a}D_a^\dag(\alpha)
\otimes D_b(\beta)(-1)^{b^\dag b}D_b^\dag(\beta)\,,
\end{equation}
where $\alpha, \beta \in {\mathbb C}$, $a$ and $b$ are mode operators and
$D_a(\alpha)=\exp\{\alpha a^\dag - \calpha a\}$ and $D_b(\beta)$ are
single-mode displacement operators.  Since the two-mode Wigner function
$W(\alpha,\beta)$ can be expressed as \cite{FOP:napoli:05}
\begin{equation}
W(\alpha,\beta) = \frac{4}{\pi^2}\, \Pi(\alpha,\beta)\,,
\end{equation}
$\Pi(\alpha,\beta)$ being the expectation value of $\hat\Pi(\alpha,\beta)$,
the violation of these inequalities is also known as nonlocality in the
phase-space. The quantity involved in such inequalities can be written as
follows
\begin{equation}\label{bell:general}
{\cal B}_{\rm DP} = \Pi(\alpha_1,\beta_1)+ \Pi(\alpha_2,\beta_1)
+ \Pi(\alpha_1,\beta_2)-\Pi(\alpha_2,\beta_2)\,,
\end{equation}
which, for local theories, satisfies $|\mathcal{B}_{\rm DP}|\le 2$.
\par
Following Ref.~\cite{bana}, one can choose a particular set of
displaced parity operators, arriving at the following combination
\cite{ips:PRA:70}
\begin{equation}
\fl
{\cal B}_{\rm DP}({\cal J}) =
\Pi(\sqrt{\cal J},-\sqrt{\cal J})+ \Pi(-3\sqrt{\cal J},-\sqrt{\cal J})
+ \Pi(\sqrt{\cal J},3\sqrt{\cal J})-\Pi(-3\sqrt{J},3\sqrt{\cal J})\,,
\label{bell:ale}
\end{equation}
which, for the TWB, gives a maximum ${\cal B}_{\rm DP} = 2.32$,
greater than the value $2.19$ obtained in Ref.~\cite{bana}. Notice that,
even in the infinite squeezing limit, the violation is never maximal, {\em
i.e.} $|\mathcal{B}_{\rm DP}| < 2\sqrt{2}$ \cite{jeong1}.
\par
In Ref.~\cite{ips:PRA:70} we studied Eq.~(\ref{bell:ale}) for both the TWB
and the IPS state in an ideal scenario, namely in the absence of
dissipation and noise; we showed that, using IPS, the maximum violation
is achieved for $\tau \to 1$ and for values of $r$ smaller than for the
TWB.
\par
\begin{figure}
\vspace{-1.5cm}
\setlength{\unitlength}{1mm}
\begin{center}
\begin{picture}(70,100)(0,0)
\put(10,0){\includegraphics[width=6cm]{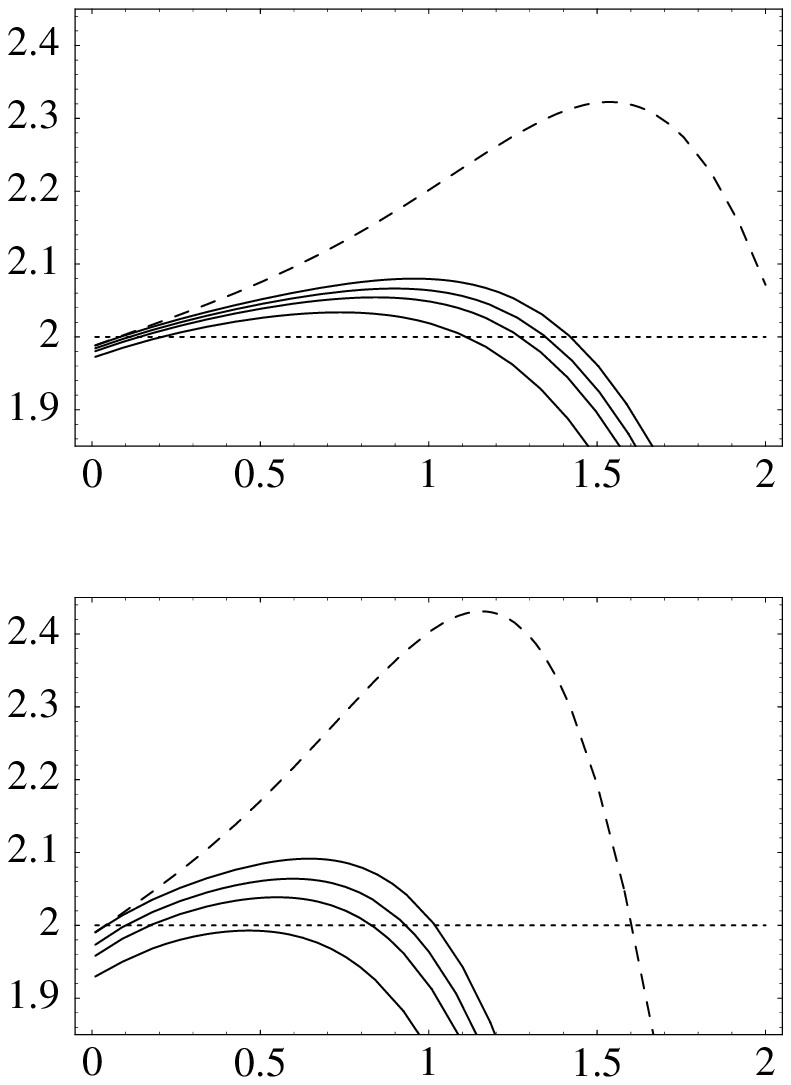}}
\put(42,42){$r$}
\put(-2,65){${\cal B}_{\rm DP}^{\rm (TWB)}$}
\put(42,0){$r$}
\put(-2,23){${\cal B}_{\rm DP}^{\rm (IPS)}$}
\end{picture}
\end{center}
\caption{Plots of the Bell parameters ${\cal B}_{\rm DP}$ for the TWB (top)
and IPS (bottom); we set ${\cal J}=1.6 \times 10^{-3}$ and $\tau = 0.9999$.
The dashed lines refer to the absence of noise ($\Gamma t = N = 0$),
whereas, for both the plot, the solid lines are ${\cal B}_{\rm DP}$ with
$\Gamma t = 0.01$ and, from top to bottom, $N=0, 0.05, 0.1,$ and $0.2$. In
the ideal case the maxima are ${\cal B}_{\rm DP}^{\rm (TWB)}=2.32$ and
${\cal B}_{\rm DP}^{\rm (IPS)}=2.43$, respectively.} \label{f:DP}
\end{figure}
Now, by means of the Eq.~(\ref{ips:wigner}) and the substitutions
(\ref{sostituzioni}), we can study how noise affects ${\cal B}_{\rm
DP}$. The results are showed in Fig.~\ref{f:DP}: as one may expect, the
overall effect of noise is to reduce the violation of the Bell's
inequality. When dissipation alone is present ($N=0$), the maximum of
violation is achieved using the IPS for values of $r$ smaller than for the
TWB, as in the ideal case. On the other hand, one can see that the presence
of thermal noise mainly affects the IPS results. In fact, for $\Gamma t =
0.01$ and $N=0.2$, one has $|{\cal B}_{\rm DP}^{\rm (TWB)}|>2$ for a range
of $r$ values, whereas $|{\cal B}_{\rm DP}^{\rm (IPS)}|$ falls below the
threshold for violation.
\par
We conclude that, considering the displaced parity test in the presence
of noise, the IPS is quite robust if the thermal noise is below a threshold
value (depending on the environmental parameters) and for small values of the
TWB parameter $r$.
\section{Nonlocality and homodyne detection} \label{s:HD}
In principle there are two approaches how to test the Bell's inequalities
for bipartite state: either one can employ some test for continuous variable
systems, such as that described in Sec.~\ref{s:DP}, or one can convert the
problem to Bell's inequalities tests on two qubits by mapping the
two modes into two-qubit systems. In this and the following Section we
will consider this latter case.
\par
The Wigner function $W_{0}^{\rm (IPS)}(\alpha,\beta)$ given in
Eq.~(\ref{ips:wigner}) is no longer positive-definite and thus
it can be  used to test the violation of Bell's
inequalities by means of  homodyne detection, {\em i.e.} measuring the
quadratures $x_{\vartheta}$ and $x_{\varphi}$ of the two IPS modes $a$ and
$b$, respectively, as proposed in Refs.~\cite{nha:PRL:93,garcia:PRL:93}.
In this case, one can dichotomize the measured quadratures assuming as
outcome $+1$ when $x \ge 0$, and $-1$ otherwise. The nonlocality of
$W_{0}^{\rm (IPS)}(\alpha,\beta)$ in ideal conditions has been studied in
Ref.~\cite{ips:PRA:70} where we also discussed the effect of the homodyne
detection efficiency $\eta_{\rm H}$.
\par
Let us now we focus our attention on
$W_{\Gamma,N}^{\rm (IPS)}(\alpha,\beta)$, namely the state produced
when the IPS process is applied to the TWB evolved through the noisy
channel. After the dichotomization of the
homodyne outputs, one obtains the following Bell parameter
\begin{equation}\label{bell:homo}
{\cal B}_{\rm HD} =
E(\vartheta_1,\varphi_1) + E(\vartheta_1,\varphi_2)
+ E(\vartheta_2,\varphi_1) - E(\vartheta_2,\varphi_2)\,,
\end{equation}
where $\vartheta_k$ and $\varphi_k$ are the phases of the two
homodyne measurements at the modes $a$ and $b$, respectively, and
\begin{equation}
E(\vartheta_h,\varphi_k) =
\int_{\mathbb{R}^2} d x_{\vartheta_h}\,d x_{\varphi_k}\,
{\rm sign}[x_{\vartheta_h}\, x_{\varphi_k}]\,
P(x_{\vartheta_h}, x_{\varphi_k})\,,
\end{equation}
$P(x_{\vartheta_h}, x_{\varphi_k})$ being the joint
probability of obtaining the two outcomes
$x_{\vartheta_h}$ and $x_{\varphi_k}$ \cite{garcia:PRL:93}. As usual,
violation of Bell's inequality is achieved when $|{\cal B}_{\rm HD}|>2$.
\par
\begin{figure}[tb]
\vspace{-1cm}
\setlength{\unitlength}{1mm}
\begin{center}
\begin{picture}(70,100)(0,0)
\put(4,0){\includegraphics[width=6.5cm]{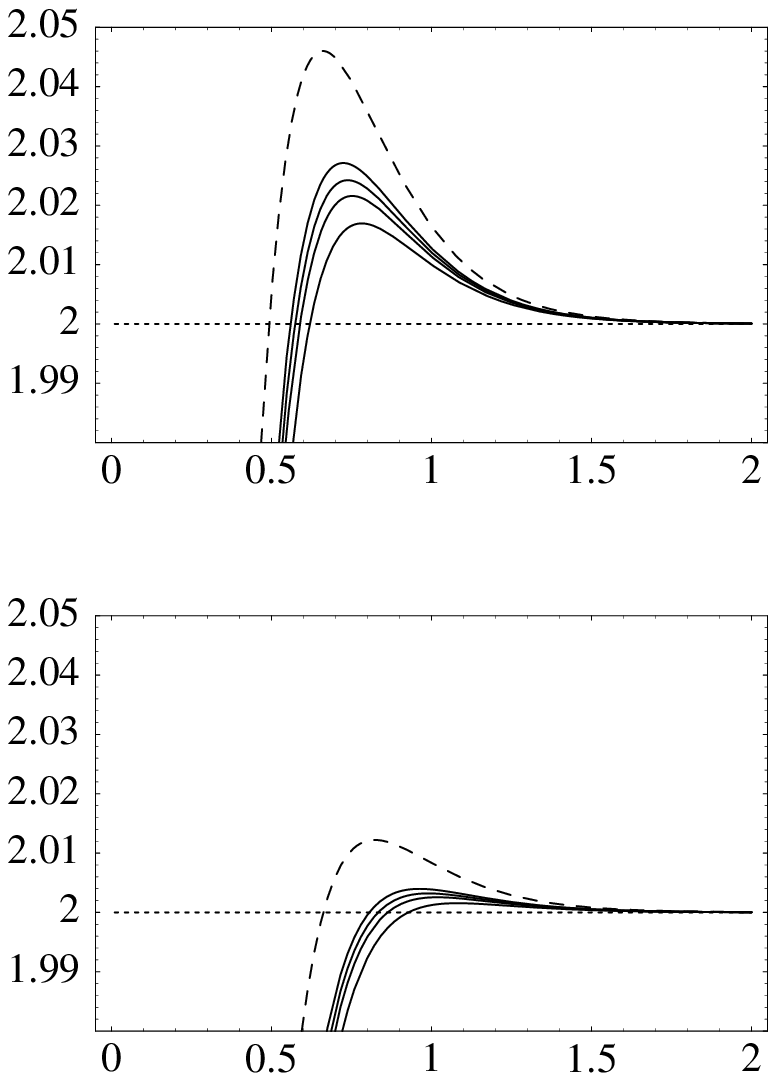}}
\put(37,45){$r$}
\put(-2,70){${\cal B}_{\rm HD}$}
\put(37,-1){$r$}
\put(-2,24.5){${\cal B}_{\rm HD}$}
\end{picture}
\end{center}
\caption{Plots of the Bell parameter ${\cal B}_{\rm HD}$ for the IPS states
for two different values of the homodyne detection efficiency: $\eta_{\rm
H} = 1$ (top), and $\eta_{\rm H}=0.9$ (bottom). We set $\tau = 0.99$. The
dashed lines refer to the absence of noise ($\Gamma t = N = 0$), whereas,
for both the plots, the solid lines are ${\cal B}_{\rm HD}$ with $\Gamma t
= 0.05$ and, from top to bottom, $N=0, 0.05, 0.1$ and $0.2$.} \label{f:HD}
\end{figure}
In Fig.~\ref{f:HD} we plot ${\cal B}_{\rm HD}$ for $\vartheta_1 = 0$,
$\vartheta_2 = \pi/2$, $\varphi_1 = -\pi/4$ and $\varphi_2 = \pi/4$: as for
the ideal case \cite{ips:PRA:70,garcia:PRL:93}, the Bell's inequality is
violated for a suitable choice of the squeezing parameter $r$. Obviously,
the presence of noise reduces the violation, but we can see that the effect
of thermal noise is not so large as in the case of the displaced parity
test addressed in Sec.~\ref{s:DP} (see Fig.~\ref{f:DP}).
\par
Notice that the high efficiencies of this kind of detectors
allow a loophole-free test of hidden variable theories
\cite{gil}, though the violations obtained are quite small.
This is due to the intrinsic information loss of the binning
process, which is used to convert the continuous homodyne data in
dichotomic results \cite{mun1}.
\section{Nonlocality and pseudospin test} \label{s:PS}
Another way to map a two-mode continuous variable system into a two-qubit
system is by means of the pseudospin test: this consists in measuring
three single-mode Hermitian operator $S_k$ satisfying the Pauli matrix algebra
$[S_h,S_k]=2i\varepsilon_{hkl}\,S_l$, $S_k^2 = {\mathbb I}$, $h,k,l=1,2,3$,
and $\varepsilon_{hkl}$ is the totally antisymmetric tensor with
$\varepsilon_{123}=+1$ \cite{filip:PRA:66,chen:PRL:88}. For the sake of
clarity, we will refer to $S_1$, $S_2$ and $S_3$ as $S_x$, $S_y$ and $S_z$,
respectively. In this way one can write the following correlation function
\begin{equation}
E({\bf a},{\bf b}) = \langle ({\bf a}\cdot{\bf S})\,
({\bf b}\cdot{\bf S})\rangle\,,
\end{equation}
where ${\bf a}$ and ${\bf b}$ are unit vectors such that
\begin{equation}
\eqalign{
{\bf a}\cdot{\bf S} &= \cos \vartheta_a\, S_z +
\sin \vartheta_a\, (e^{i \varphi_a} S_{-} + e^{-i \varphi_a} S_{+})\,,\\
{\bf b}\cdot{\bf S} &= \cos \vartheta_b\, S_z +
\sin \vartheta_b\, (e^{i \varphi_b} S_{-} + e^{-i \varphi_b} S_{+})\,,
}
\end{equation}
with $S_{\pm} = \frac12 (S_x \pm S_y)$. In the following, without loss of
generality, we set $\varphi_k = 0$. Finally, the Bell parameter reads
\begin{equation}\label{bell:PS}
{\cal B}_{\rm PS} = E({\bf a}_1,{\bf b}_1)+E({\bf a}_1,{\bf b}_2)
+E({\bf a}_2,{\bf b}_1)-E({\bf a}_2,{\bf b}_2)\,,
\end{equation}
corresponding to the CHSH Bell's inequality $|{\cal B}_{\rm PS}|\le 2$. In
order to study Eq.~(\ref{bell:PS}) we should choose a specific
representation of the pseudospin operators; note that, as pointed out in
Refs.~\cite{revzen, ferraro:3:nonloc}, the violation of Bell inequalities
for continuous variable systems depends, besides on the orientational
parameters, on the chosen representation, since different $S_k$ leads to
different expectation values of ${\cal B}_{\rm PS}$. Here we consider the
pseudospin operators corresponding to the Wigner functions \cite{revzen}
\begin{eqnarray}
W_x(\alpha)&=\frac{1}{\pi}\,{\rm sign}\big[\Re{\rm e}[\alpha]\big]\,,\quad
W_z(\alpha)= -\frac{1}{2}\,\delta^{(2)}(\alpha)\,,\label{PS:W:xz}\\
&W_y(\alpha)=-\frac{1}{2\pi}\, \delta\big(\Re{\rm e}[\alpha] \big)\,
{\cal P} \frac{1}{\Im{\rm m}[\alpha]}\,,
\end{eqnarray}
where ${\cal P}$ denotes the Cauchy's principal value. Thanks to
(\ref{PS:W:xz}) one obtains
\begin{equation}
E_{\rm TWB}({\bf a},{\bf b}) = \cos\vartheta_a \cos\vartheta_b 
+ \frac{2\sin\vartheta_a \sin\vartheta_b}{\pi}\,
\arctan\big[ \sinh(2r) \big]\,,
\end{equation}
for the TWB, and, for the IPS,
\begin{equation}
\fl
E_{\rm IPS}({\bf a},{\bf b}) = 
\sum_{k=1}^4 \frac{{\cal C}_k(r,\tau)}{p_{11}(r,\tau)}
\Bigg[
\frac{\cos\vartheta_a \cos\vartheta_b}{4}
+ \frac{2 \sin\vartheta_a \sin\vartheta_b}{\pi{\cal A}_k}\,
\arctan\left( \frac{2 \wtB_0 \tau + h_k}{\sqrt{{\cal A}_k}} \right)
\Bigg]
\end{equation}
where $ {\cal A}_k=(b-f_k)(b-g_k)-(2 \wtB_0 \tau + h_k)^2$,
and all the other quantities have been defined in Sec.~\ref{s:IPS}.
\par
\begin{figure}[tb]
\vspace{-1cm}
\setlength{\unitlength}{1mm}
\begin{center}
\begin{picture}(70,50)(0,0)
\put(4,0){\includegraphics[width=6.5cm]{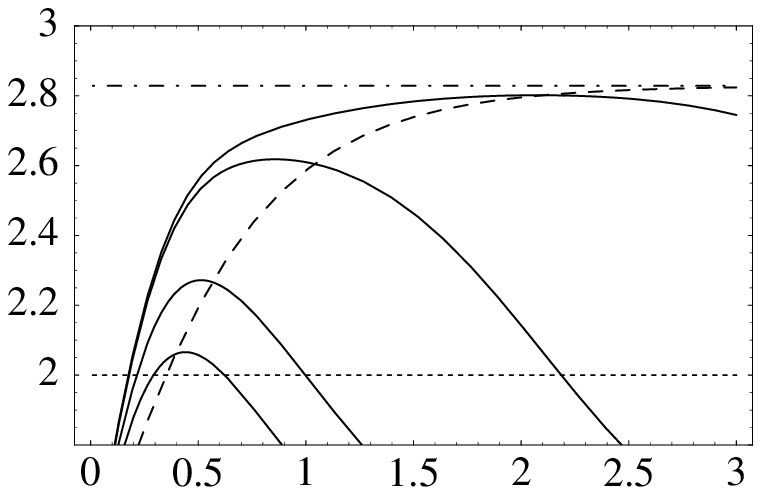}}
\put(40,-3){$r$}
\put(-2,24.5){${\cal B}_{\rm PS}$}
\end{picture}
\end{center}
\caption{Plots of the Bell parameter ${\cal B}_{\rm PS}$ in ideal case
($\Gamma t = N = 0$): the dashed line refers to the TWB, whereas the solid
lines refer to the IPS with, from top to bottom, $\tau = 0.9999, 0.99,
0.9$, and $0.8$. There is a threshold value for $r$ below which IPS gives a
higher violation than TWB.  Note that there is also a region of small
values of $r$ for which the IPS state violates the Bell's inequality while
the TWB does not. The dash dotted line is the maximal violation value
$2\sqrt{2}$.} \label{f:PS:id}
\end{figure}
In Fig.~\ref{f:PS:id} we plot ${\cal B}_{\rm PS}$ for the TWB and IPS in
the ideal case, namely in the absence of dissipation and thermal noise. For
all the Figures we set $\vartheta_{a_1}=0$, $\vartheta_{a_2}=\pi/2$, and
$\vartheta_{b_1}=-\vartheta_{b_2}=\pi/4$. As
usual the IPS leads to better results for small values of $r$. Whereas
${\cal B}_{\rm PS}^{\rm (TWB)} \to 2\sqrt{2}$ as $r\to \infty$,
${\cal B}_{\rm PS}^{\rm (IPS)}$ has a maximum and, then, falls below the
threshold $2$ as $r$ increases. It is interesting to note that there is a
region of small values of $r$ for which  ${\cal B}_{\rm PS}^{\rm (TWB)}\le
2 < {\cal B}_{\rm PS}^{\rm (IPS)}$, {\em i.e.} the IPS process can increases
the nonlocal properties of a TWB which does not violates the Bell's
inequality for the pseudospin test, in such a way that the resulting state
violates it. This fact is also present in the case of the displaced parity
test described in Sec.~\ref{s:DP}, but using the pseudospin test the effect
is enhanced. Notice that the maximum violations for the IPS occur for a
range of values $r$ experimentally achievable.
\par
\begin{figure}[tb]
\vspace{-1cm}
\setlength{\unitlength}{1mm}
\begin{center}
\begin{picture}(70,50)(0,0)
\put(4,0){\includegraphics[width=6.5cm]{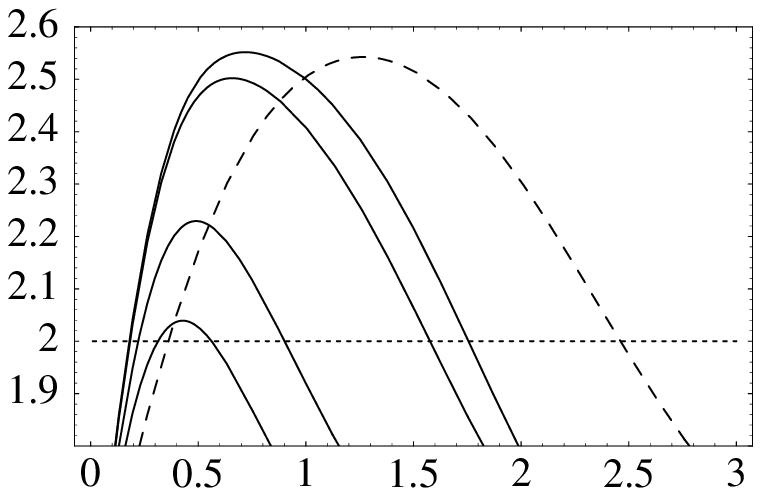}}
\put(40,-3){$r$}
\put(-2,24.5){${\cal B}_{\rm PS}$}
\end{picture}
\end{center}
\caption{Plots of the Bell parameter ${\cal B}_{\rm PS}$ for $\Gamma t =
0.01$: the dashed line refers to the TWB, whereas the solid lines refer to
the IPS with, from top to bottom, $\tau = 0.9999, 0.99, 0.9$, and $0.8$.
The same comments as in Fig.~\ref{f:PS:id} still hold.} \label{f:PS:tau}
\end{figure}
\begin{figure}[tb]
\vspace{-1cm}
\setlength{\unitlength}{1mm}
\begin{center}
\begin{picture}(70,50)(0,0)
\put(4,0){\includegraphics[width=6.5cm]{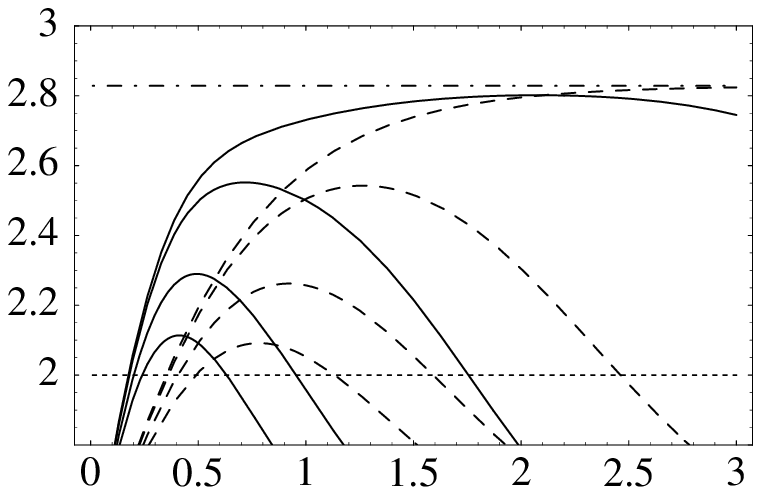}}
\put(40,-3){$r$}
\put(-2,24.5){${\cal B}_{\rm PS}$}
\end{picture}
\end{center}
\caption{Plots of the Bell parameter ${\cal B}_{\rm PS}$ for different
values of $\Gamma t$ and in the absence of thermal noise ($N = 0$): the
dashed lines refer to the TWB, whereas the solid ones refer to the IPS with
$\tau = 0.9999$; for both the TWB and IPS we set, from top to bottom,
$\Gamma t = 0, 0.01, 0.05$, and $0.1$. The dash dotted line is the maximal
violation value $2\sqrt{2}$.} \label{f:PS:gamma}
\end{figure}
\begin{figure}[tb]
\vspace{-1cm}
\setlength{\unitlength}{1mm}
\begin{center}
\begin{picture}(70,50)(0,0)
\put(4,0){\includegraphics[width=6.5cm]{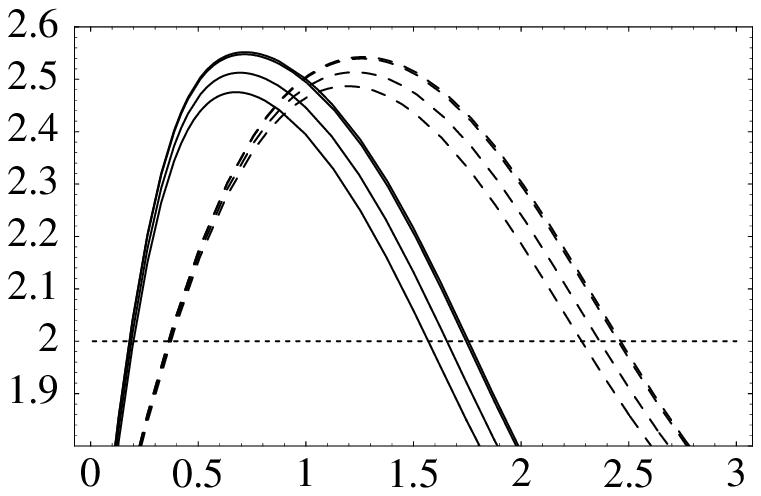}}
\put(40,-3){$r$}
\put(-2,24.5){${\cal B}_{\rm PS}$}
\end{picture}
\end{center}
\caption{Plots of the Bell parameter ${\cal B}_{\rm PS}$ for $\Gamma t =
0.01$ and different values $N = 0$: the dashed lines refer to the TWB,
whereas the solid ones refer to the IPS with $\tau = 0.9999$; for both the
TWB and IPS we set, from top to bottom, $N = 0, 0.01, 0.1$, and $0.2$.}
\label{f:PS:th}
\end{figure}

In Fig.~\ref{f:PS:tau} we consider the presence of the dissipation alone
and vary $\tau$. We can see that IPS is effective also when the
effective transmissivity $\tau$ is not very high.
We take into account the effect of dissipation and thermal noise
in Figs.~\ref{f:PS:gamma}, and \ref{f:PS:th}: we can conclude that
IPS is quite robust with respect to this sources of noise and, moreover,
one can think of employing IPS as a useful resource in order to reduce the
effect of noise.
\section{Concluding remarks} \label{s:remarks}
We have addressed three different nonlocality tests, namely, displaced
parity, homodyne detection and pseudospin test, on TWB and IPS in the
presence of noise. We have shown that the IPS process on TWB enhances
nonlocality not only in ideal cases, but also when noise (dissipation and
thermal noise) affects the propagation. As in the ideal situation, the
enhancement is achieved when the TWB energy is not too high (small
squeezing parameter $r$), depending on the environmental parameters.
Moreover, in the case of the pseudospin test, we have seen that there is a
region of small $r$ for which the TWB itself does not violates the Bell's
inequality, wheres after the IPS process it does.
\par
Finally, we mention that the enhanced nonlocality also in the presence of
noise makes the IPS states useful resources for continuous variable quantum
information processing.

\ack
Stimulating and useful discussions with M.~S.~Kim, A.~Ferraro and
A.~R.~Rossi are gratefully acknowledged.

\end{document}